\documentclass{article}%
\usepackage{amsmath}
\usepackage{geometry}%
\usepackage{caption, booktabs}%
\usepackage{color}
\usepackage{float}%
\usepackage{amsfonts}%
\usepackage{amssymb}%
\usepackage{graphicx}
\providecommand{\U}[1]{\protect\rule{.1in}{.1in}}

\begin{document}
\[%
\begin{tabular}
[c]{c}%
$\underset{}{}${\LARGE Sequentially estimating the dynamic contact angle of}$\underset{}{}$\\
$\underset{}{}${\LARGE sessile saliva droplets in view of SARS-CoV-2}$\underset{}{}$\\
\end{tabular}
\]

\[%
\begin{tabular}
[c]{c}%
$\text{Sudeep R. Bapat}$\\
$\underset{}{}$Indian Institute of Management, Indore, India$\underset{}{}$\\
E-mail: sudeepb@iimidr.ac.in
\end{tabular}
\]%
\[
\]

\noindent\textbf{Abstract: }Estimating the contact angle of a virus infected saliva droplet is seen to be an important area of research as it presents an idea about the drying time of the respective droplet and in turn of the growth of the underlying pandemic. In this paper we extend the data presented by Balusamy, Banerjee and Sahu [``Lifetime of sessile saliva droplets in the context of SARS-CoV-2," Int. J. Heat Mass Transf. \textbf{123}, 105178 (2021)], where the contact angles are fitted using a newly proposed \textit{half-circular wrapped-exponential} model, and a sequential confidence interval estimation approach is established which largely reduces both time and cost with regards to data collection.

\vspace{0.08in}\vspace{0.08in}


\vspace{0.05in}\vspace{0.05in}

\bigskip

\noindent\textbf{1. INTRODUCTION}

\bigskip

\noindent SARS-CoV-2 (virus which causes COVID-19) has severely impacted more than 200 countries worldwide, with over 180 million cases until the end of June, 2021. The span of this virus was so fast and devastating that the World Health Organization declared the outbreak as a Public Health Emergency of International Concern on 30 January, 2020, whereas a global pandemic on 11 March, 2020. Spreading of such respiratory diseases is largely caused due to respiratory droplets of saliva (of an infected person) during coughing, sneezing or even moist speaking. A recent reference paper in this regard is by Balusamy et al. [1]. Understanding the lifetime of such droplets is hence an important area of research, which could be handled by studying the fluid dynamics of such droplets in air. One may refer to Mittal et al. [2] who analyze the flow-physics of virus laden respiratory droplets, or Bhardwaj and Aggarwal [3] who analyze the likelihood of survival of a virus laden droplet on a solid surface. Further, it has been studied that such respiratory droplets have a tendency to increase their lifetime on coming in contact with a surface based on its properties. Vejerano and Marr [4] studied the physico-chemical characteristics of evaporating respiratory fluid droplets and found out that a typical saliva droplet also contains NaCl, mucin (protein) and a certain surfactant in fixed amounts. In addition to the droplet composition, the evaporation rate of a droplet also depends on environmental conditions and factors such as temperature, relative humidity, droplet volume and the contact angle which the droplet makes with the surface. A specific analysis was carried out in [3] where the authors examined the drying time of a deposited droplet in two different temperatures namely, $25^{\circ}$C and $40^{\circ}$C which represent an air-conditioned room and a summer afternoon respectively. The contact angle and humidity were set at $30^{\circ}$ and $50\%$. Studying the drying time of a droplet plays an important role as it well related to the the survival of the droplet and in turn to the growth of the pandemic. Chaudhuri et al. [5] tested this hypothesis using suspended droplets in air, whereas [3] compared the growth of infection with the drying time in different cities globally. They verified that for a $5$ nL droplet, a higher drying time corresponds to a higher pandemic growth rate. Hence, when a droplet evaporates slowly, the chance of the survival of the virus is enhanced.

Specifically, the initial contact angle, which measures the angle that a droplet makes with the surface plays a big role in determining the lifetime of it. Different contact angles are predominant with different surfaces i.e., droplets on glass, wood, stainless steel, cotton or the touchscreen of a smartphone tend to make angles varying from $5^{\circ}$ to $95^{\circ}$. It is also intuitive that a contact angle cannot exceed $180^{\circ}$. Figure 1 contains pictorial representations of two different droplets making different angles with the surface. The left image shows a water droplet on cloth, making a high contact angle due to the hydrophobic property of the cloth. Whereas the image on the right shows a water droplet on a lotus leaf, again making a high contact angle. Both the images are borrowed from Wikipedia under the license CC BY-SA 3.0.

\[%
\begin{tabular*}
{\textwidth}[c]{@{\extracolsep{\fill}}cc}%
$\includegraphics[width = .43\textwidth]{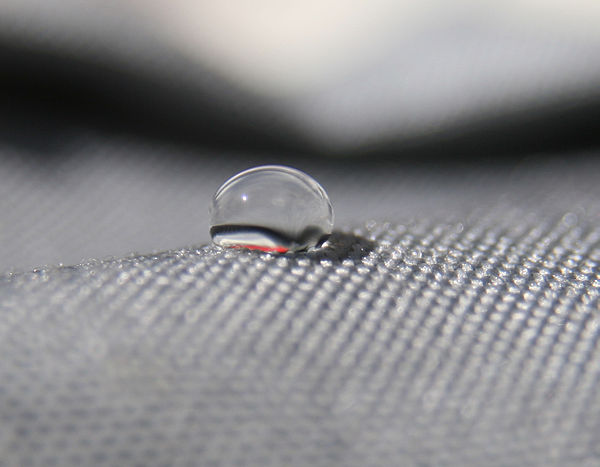}$ & $\includegraphics[width = .45\textwidth]{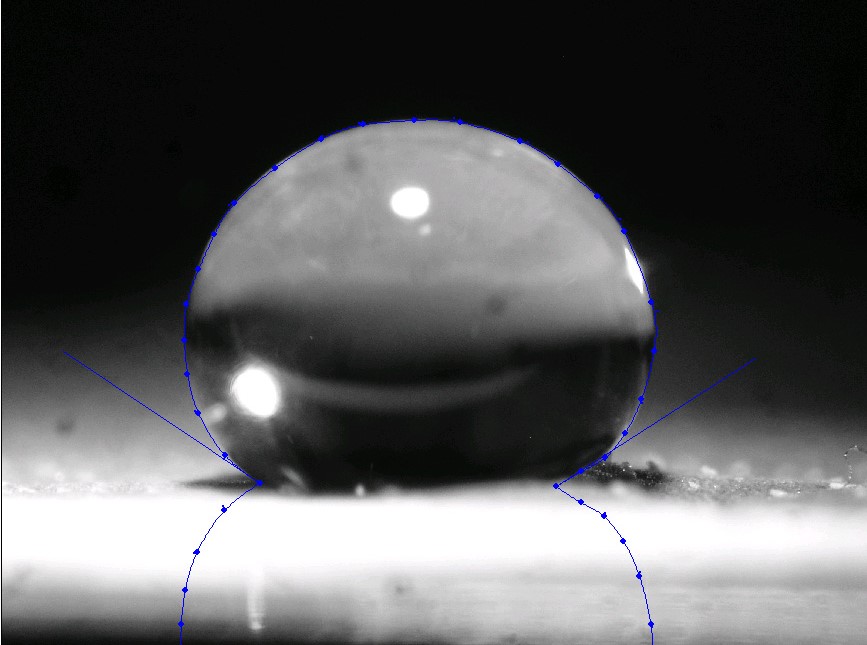}$  \\
(a) water drop on cloth & (b) water drop on a lotus leaf \\ \\
\multicolumn{2}{c}{\textbf{Figure 1}. Water droplets making contact angles greater than $90^{\circ}$ on two different surfaces. Both the}\\
\multicolumn{2}{c}{images are borrowed from \textit{Wikipedia} under the license CC BY-SA 3.0}
\end{tabular*}
\]

A dynamic contact angle is the one which is measured as the droplet changes its size as it moves quickly over the surface. One may again refer to [3] or [1] for more details. However it is also true that measuring such contact angles (initial or dynamic) involves a lot of struggle and cost, as it has to be carried out using heavy apparatus. Some of the existing methods for contact angle determination include the \textit{sessile droplet method}, where the angle is measured using a ``contact angle goniometer", the \textit{pendant drop method} which is used to measure angles for pendant drops, the \textit{dynamic sessile drop method} which is similar to a sessile drop method but requires the drop to be modified or a \textit{single-fiber meniscus method} where the shape of the meniscus on the fiber is directly imaged through a high resolution camera. One may refer to Albert et al. [6] for an overview of other techniques. Hence to estimate a dynamic contact angle of a droplet, a reduction in the number of observations required to carry out the estimation is highly beneficial. In this paper, we thus introduce a sequential estimation technique which is outlined in Section 2.

Now since the aim of this paper is to estimate a certain contact angle, it makes more sense to apply a circular model rather than a usual linear one on the concerned data. Literature on such models is vast and ever expanding. A few other examples where a circular model is appropriate involve orientations of the major axis of termite mounds, the angles of slope of different sedimentary layers of an exposed rock-face or the walking directions of long legged desert ants etc. In all these examples, the observations are either certain directions, or angles measured in degrees or radians. Such observations are often measured either clockwise or counter-clockwise from some reference direction, usually called as the \textit{zero direction}. Over years, a usual technique to design new circular distributions is to wrap a linear distribution over a full circle. However as seen before, since the contact angles of any droplet is necessarily less than $180^{\circ}$, an adjusted model which is capable of taking values only on half-a-circle seems more appropriate. In this context, we introduce a new model called as the \textit{half-circular wrapped-exponential distribution} to model our data. In general, a few notable books covering circular models which one can refer to are by Mardia and Jupp [7], Rao and Sengupta [8], Rao and Girija [9] or Ley and Verdebout [10], among others. 

\bigskip

\noindent\textbf{2. Data modeling and analysis}

\bigskip

The particular dataset analyzed for this experiment is a pseudo dataset which is an extended version of the one borrowed from [1] and consists of the temporal variations of the dynamic contact angles in degrees (simply called as contact angles from now on) of the droplet normalized with the initial contact angle, $\theta/\theta_0$. The particular setting used for this experiment is as follows: the relative humidity (RH) is controlled at $50\%$, the initial droplet volume $(V_0)$ is $10$ nL, the molality of the saliva $(M)$ is $0.154$ mol/kg, temperature $(T)$ is $30^{\circ}$, the surfactant parameter $(\Psi)$ is $10$ and the initial contact angle $(\theta_0)$ is $50^{\circ}$. One may refer to Figure 2a in [1] for a pictorial description of the dataset. As there was not an access to the actual observations, we adopted the following approach: for brevity alone, we only focused on the curve representing RH$=50\%$. Using an online tool, we extracted the $(x,y)$ coordinates for each of its 20 observations. We converted these normalized contact angles to actual contact angles $(\theta)$ and finally translated those into radians. Table 1 lists all these observations for convenience. Now, to include more observations in the analysis, we first assumed a linear relationship between ``time" and ``contact angles" (CA), fitted several polynomial regression models and picked the following third-order model which fitted better with a $R^2$ value of $0.9613$. 

\begin{equation}
CA = 0.985-8.45\times 10^{-3} \,time+2.34\times 10^{-5}\, time^2-2.05\times 10^{-8} \,time^3
\end{equation} 

Figure 2 contains a scatterplot of the raw data (1a) and the fitted polynomial regression model superimposed on it (1b).

\[%
\begin{tabular}
[c]{cccccc}%
\multicolumn{6}{c}{\textbf{Table 1. }Extracted dataset containing the temporal variations}\\ 
\multicolumn{6}{c}{of the contact angles (in radians)}\\ \\
$\underset{}{\overset{}{\text{Time (sec)}}}$ & CA & Time (sec) & CA &
Time (sec) & CA\\\hline
$10$ & $0.811$ & $88.75$ & $0.379$ & $287.5$ & $0.034$\\
$25$ & $0.794$ & $100$ & $0.261$ & $325$ & $0.031$\\
$55$ & $0.689$ & $118.75$ & $0.218$ & $381.25$ & $0.028$\\
$58.75$ & $0.654$ & $137.5$ & $0.157$ & $437.5$ & $0.026$\\
$66.25$ & $0.593$ & $175$ & $0.109$ & $493.75$ & $0.023$\\
$77.25$ & $0.471$ & $212.5$ & $0.052$ & $550$ & $0.020$\\
$83.15$ & $0.436$ & $250$ & $0.035$ & $$ & $$\\\hline
\end{tabular}
\]\\

\[%
\begin{tabular*}
{\textwidth}[c]{@{\extracolsep{\fill}}cc}%
$\includegraphics[width = .5\textwidth]{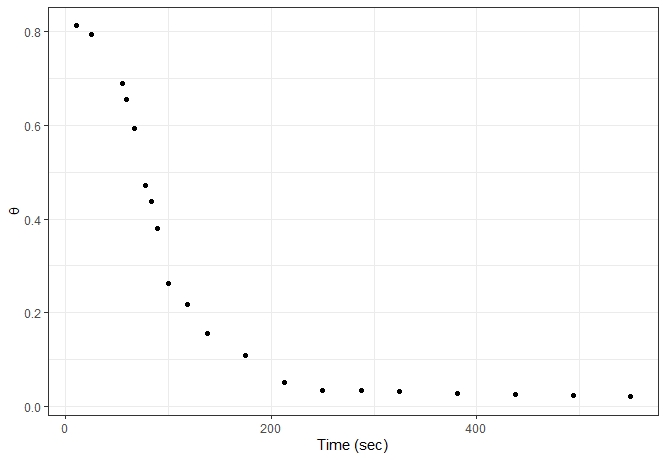}$ & $\includegraphics[width = .5\textwidth]{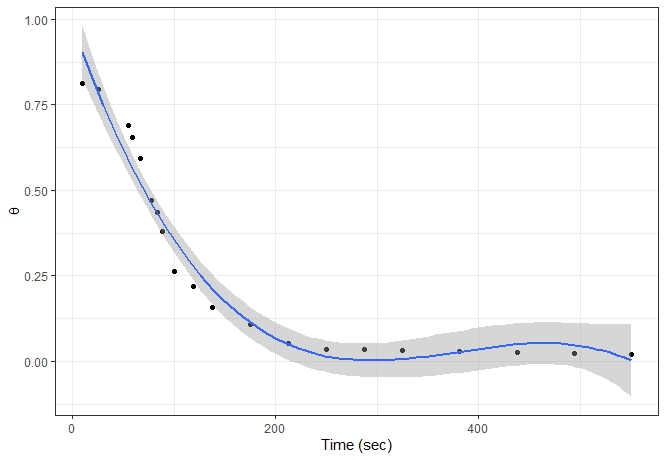}$  \\
(a) plot of the raw data & (b) superimposed polynomial model \\ \\
\multicolumn{2}{c}{\textbf{Figure 2}. Temporal variations of the contact angles}
\end{tabular*}
\]

We then assumed a vector of times ranging over $5-300$ seconds with a jump of $1$ second in between, and predicted the contact angles according to the above model. Thus, our final pseudo dataset consists of $296$ observations according to our construction.

\bigskip

\noindent\textbf{2.1. A half-circular wrapped-exponential model for the contact angles}

\bigskip

For a start, Figure 3 shows a pictorial distribution of our pseudo data placed around a circle. Purposefully, we have stacked the closely lying observations for a better visualization and as one can observe, all the observations lie entirely between $0$ and $\pi/2$ radians. As seen before, wrapping a linear density over a circle is a suitable choice to model such observations. In this case, since the linear curve seen in Figure 2 shows an exponential decline, it makes sense to choose some of the lifetime distributions and wrapping them around a circle. Now as discussed before, since any contact angle of a droplet is always less than $\pi$ radians it makes more sense to fit a distribution which takes values only on a semicircle. In literature, not many such distributions have been proposed. One such example is of a half-circular distribution which was introduced by Rambli et al. [11], who converted a Gamma distribution to a half-circular one and fitted it to the angle which measures the posterior corneal curvature of an eye. In a similar spirit, we now introduce a \textit{half-circular wrapped-exponential} (HCWE) distribution with parameter $\lambda$. An intuitive construction is through the following transformation: $X_w=X(\text{mod}\, \pi)$, where $X$ is a linear exponential random variable with pdf $f(x)=\lambda e^{-\lambda x}, \, x>0, \lambda>0$. Interestingly, another easy construction is to simply truncate $X$ over $[0, \pi)$. Its pdf, cdf and characteristic functions are as follows,
\begin{equation}
f_w(\theta)=\frac{\lambda e^{-\lambda \theta}}{1-e^{-\pi \lambda}}, \hspace{.3 cm} \theta \in [0, \pi)
\end{equation}

\begin{equation}
F_w(\theta)=\frac{1-e^{-\lambda \theta}}{1-e^{-\pi \lambda}}, \hspace{.3 cm} \theta \in [0, \pi)
\end{equation}

\begin{equation}
\phi_p=\frac{1}{1-ip/\lambda}, \hspace{.3 cm} p=0, \pm 1, \pm 2,...
\end{equation}
Consequently, the mean direction happens to be,
\begin{equation}
\mu_0=\tan^{-1}\frac{1}{\lambda}, \hspace{.3 cm} \lambda>0
\end{equation}
Now for a comparison, we tried to fit several other wrapped distributions to the data namely, the wrapped-exponential by Jammalamadaka and Kozubowski [12], transmuted wrapped-exponential by Yilmaz and Bi\c cer [13] and wrapped-Lindley by Joshi and Jose [14]. For completeness, we also fit a von-Mises distribution which is one of the widely used circular models. Table 2 contains the log-likelihood values and the AICs for these five models. As one can observe, the half-circular wrapped-exponential model fits better than the others. It is also seen to be a significant fit with a p-value of $0.18$ using the Kolmogorov-Smirnov test, and the estimated $\lambda$ value equals $3.69$. On using Eq. (5), the estimated mean direction equals $0.2646$ radians. Figure 4 contains a set of goodness of fit plots for the HCWE$(\lambda)$ distribution. All these fits and plots were carried out using the $``circular"$ and $``fitdistrplus"$ packages in $R$.

\begin{center}
\includegraphics[width = .5\textwidth]{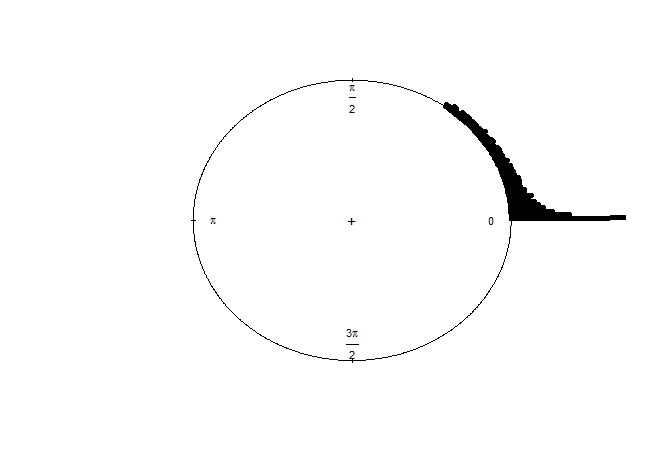}\\
{\textbf{Figure 3}. Raw circular plot of the pseudo data }
\end{center}

\begin{center}
\includegraphics[width = .7\textwidth]{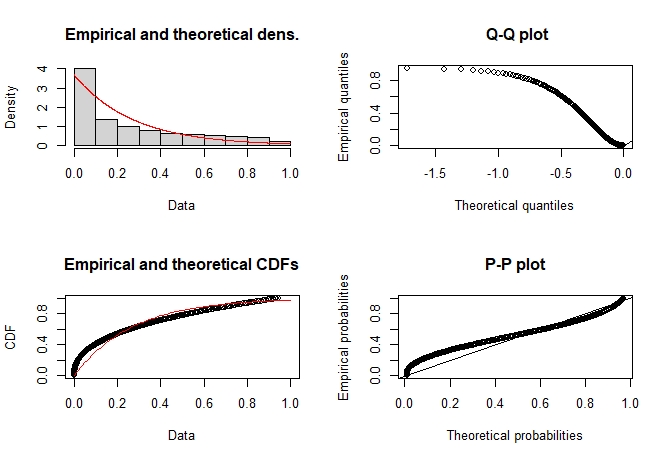}\\
{\textbf{Figure 4}. Goodness of fit plots for the HCWE$(\lambda)$ density on the pseudo data}
\end{center}

Now, since in practice the value of $\lambda$ will be unknown, we develop a sequential fixed-width confidence interval to estimate $\lambda$ which in turn will give us an estimate for the mean direction $\mu_0$ of the contact angle, which will give us a fair idea about the drying time of the droplet.

\[%
\begin{tabular}
[c]{ccc}%
\multicolumn{3}{c}{\textbf{Table 2. }Comparing model fits to the pseudo data}\\ \\
$\underset{}{\overset{}{\text{Model}}}$ & Log-likelihood & AIC\\\hline
von-mises & $-36.72$ & $77.44$\\
wrapped-exponential & $90.75$ & $-179.50$ \\
\textbf{half-circular wrapped-exponential} & $\textbf{92.56}$ & $\textbf{-181.52}$\\
transmuted wrapped-exponential & $-7.01$ & $18.02$ \\
wrapped-Lindley & $89.94$ & $-177.88$ \\
\end{tabular}
\]\\

\bigskip

\noindent\textbf{2.2. A sequential fixed-width confidence interval}

\bigskip

In general, a sequential rule consists of identifying a stopping variable, which determines the optimal sample size to be used in the experiment. This technique largely reduces the number of observations needed for the inference part, which proves to be beneficial as it reduces both time and cost. Literature on sequential estimation methodologies is vast and still being explored. In particular, a few recent works aimed at finding appropriate confidence intervals include, Banerjee and Mukhopadhyay [15], who developed a general sequential fixed-accuracy confidence interval, Mukhopadhyay and Banerjee [16], who looked at constructing bounded length intervals, Bapat [17, 18] who constructed fixed-accuracy intervals for parameters under an inverse Gaussian and bivariate exponential models or Khalifeh et al. [19], who derived fixed-accuracy intervals for the reliability parameter of an exponential distribution.

To summarize, a fixed-width interval (FWI) aims at simultaneously controlling the width of the interval (say, $d)$ and the confidence limit $(1-\alpha)$. Such an interval is clearly symmetric around the parameter. It turns out that there does not exist any fixed sample size procedure to tackle this problem and one has to resort to a sequential setup. However, a certain drawback of this method is: even though a parameter is entirely positive, the lower bound of a FWI can assume negative values. A fix to this is to construct a fixed-accuracy interval (FAI), which assumes a fixed-accuracy value (say, $d)$. A FAI happens to be symmetric around $\log$ of the parameter. An introductory paper to this approach is Banerjee and Mukhopadhyay [20]. Even in this case it may happen, that if the parameter space is bounded (say from above by $U$), a FAI may contain bounds which cross  $U$. Hence, [15] came up with a bounded-length fixed-accuracy interval (BLFAI) as a fix. In our case, we aim at constructing a fixed-width interval as outlined next.

Let $\theta_1, \theta_2,...$ be the dynamic contact angles of a droplet, measured using a suitable technique. Then, for some pre-fixed width $d$, a confidence interval of $\lambda$ takes the following form,
\begin{equation}
I_n =\left\{ \lambda: \lambda \in \left[\hat{\lambda}_n-d, \hat{\lambda}_n+d \right]\right\},
\end{equation}
where $\hat{\lambda}_n$ is the MLE of $\lambda$, which is consistent and asymptotically normal with the following representation,
\begin{equation}
\sqrt{n}(\hat{\lambda}_n-\lambda) \xrightarrow{D} N\left( 0, \sigma^2_{\hat{\lambda}_n}\right),
\end{equation}
where $\sigma^2_{\hat{\lambda}_n}$ is the variance of the MLE and $\xrightarrow{D}$ stands for convergence in distribution. Now, for $I_n$ to include $\lambda$ with a pre-fixed coverage probability $1-\alpha$, the required fixed sample size can be found out as follows,
\begin{equation*}
P\left( \hat{\lambda}_n -d \leq \lambda \leq \hat{\lambda}_n+d \right) = 1-\alpha
\end{equation*}
\begin{equation}
\Rightarrow n^* \equiv n_d^* =\left( \frac{z_{\alpha/2}}{d}\right)^2 \sigma^2_{\hat{\lambda}_n},
\end{equation}
where $z_{\alpha/2}$ is the upper $100(\alpha/2)\%$ point of a standard normal distribution. Since $n^*$ is an unknown quantity, we now propose the following sequential methodology:

We first fix an integer $m(>1)$ often called as the ``pilot sample size" and obtain a pilot sample $\theta_1, \theta_2,...,\theta_m$ from a HCWE$(\lambda)$ density as given in (5). We then aim to collect an additional observation at every stage, until sampling is terminated according to the following stopping rule:
\begin{equation}
N=\inf\left\{ n \geq m: n \geq \left( \frac{z_{\alpha/2}}{d}\right)^2 \hat{\sigma}^2_{\hat{\lambda}_n} \right\},
\end{equation} 
where $\hat{\sigma}^2_{\hat{\lambda}_n}$ is the estimated variance of the MLE. We then have a final set of observations $\theta_1, \theta_2,...,\theta_N$ and will estimate $\lambda$ using the interval,
\begin{equation}
I_N = \left[ \hat{\lambda}_N-d, \hat{\lambda}_N+d\right] = \left[ L_N, U_N\right] \, \text{(say)}.
\end{equation}
The stopping variable $N$ from (12) follows properties such as asymptotic first-order efficiency and asymptotic consistency. We leave out the proofs for brevity. One may refer to Theorem 3.1 of [17]. Finally, we estimate the mean direction $\mu_0$ using an interval,
\begin{equation}
J_N = \left[ \tan^{-1}\frac{1}{U_N}, \, \, \tan^{-1}\frac{1}{L_N}\right].
\end{equation}
We now outline a stepwise procedure to tackle a practical problem through the above methodology.
\begin{itemize}
\item[\textbf{Step 1:}] For a certain specific liquid droplet, observe the contact angles over equally spaced time intervals and note down the first $m$ angles $(\theta_1, \theta_2,...,\theta_m)$ over the first $m$ time points $t_1,t_2,...,t_m.$
\item[\textbf{Step 2:}] After $t_m$, collect observations (i.e. observe contact angles) one-at-a-time according to the stopping rule given in (9).
\item[\textbf{Step 3:}] Once the stopping rule is executed, observe the value of $N$, find out an interval for $\lambda$ as per (10) and ultimately find a subsequent interval for the mean direction $\mu_0$ according to (11).  
\item[\textbf{Step 4:}] Using the interval for $\mu_0$, find out a rough interval for the average drying time of the droplet by predicting using the following inverted polynomial regression model $(R^2 = 0.98)$ (i.e. by assuming ``time" as the response and ``contact angle" as the predictor. 

\begin{equation}
time = 266.96-872.293 \,CA+1329.892\, CA^2-763.05 \,CA^3
\end{equation}
\end{itemize}
Hence, for our complete pseudo data, $\hat{\lambda}_n=3.69, \hat{\mu}_0=0.2646$ and the estimated drying time equals $115.13$ seconds. We now apply the above procedure to our observed pseudo data with a small adjustment: we first randomize the entire data, sample $250$ observations, and sort them. This kind of an approach gives a good representation of the actual data in every simulation. We consider several fixed values of $d$ ranging from $0.05$ to $0.6$ over roughly equally spaced intervals. We fix the pilot sample size $m=5$ and the significance level $\alpha=0.05.$ After implementing the sequential rule (9) with a particular choice of $d,$ we obtain the confidence interval for $\lambda$ and in turn report the interval for $\mu_0$, and finally an interval for the average drying time of the droplet. Since the procedure has to be solved analytically, all the analyses were carried out again using the $``fitdistrplus"$ package in $R$.

\[%
\begin{tabular}
[c]{ccccc}%
\multicolumn{5}{c}{\textbf{Table 3. }Analysis of the CA data using
purely}\\
\multicolumn{5}{c}{sequential methodology (12) with $m=5,$ $\alpha=0.05$%
}\\\hline
$\underset{}{\overset{}{d}}$ & $N$ & CI for $\lambda$ & CI for $\mu_0$ & $\underset
{}{\overset{}{\text{CI for Drying time (s)}}}$\\\hline
$\overset{}{0.05}$ & $214$ & $(2.65, 2.75)$  & $(0.34, 0.36)$ & $(89.69, 94.13)$\\
$0.1$ & $176$ & $(2.19, 2.39)$  & $(0.39, 0.42)$ & $(78.66, 83.78)$\\
$0.2$ & $141$ & $(1.74, 2.14)$  & $(0.43, 0.51)$ & $(66.78, 77.11)$\\
$0.3$ & $132$ & $(1.56, 2.16)$  & $(0.43, 0.56)$ & $(61.53, 77.11)$\\
$0.4$ & $126$ & $(1.29, 2.09)$  & $(0.44, 0.65)$ & $(52.30, 75.62)$\\
$0.5$ & $112$ & $(1.19, 2.19)$  & $(0.42, 0.69)$ & $(47.57, 78.66)$\\
$0.6$ & $101$ & $(0.99, 2.19) $ &  $(0.42, 0.78)$ & $(33.57, 78.66)$\\\hline
\end{tabular}
\]\\
A few take away points from Table 3 are: as one goes on increasing $d$, naturally, the width of the desired interval increases and as a result, less number of observations are required to achieve a confidence level of $\alpha$ $(0.05$ in this case). Also, for increasing $N$, the intervals for the drying time also increase and are seen to approach the actual estimated drying time of $115.13$ seconds. But of course, a larger sample size comes with a cost and hence one needs to strike a proper balance.

\bigskip
\noindent\textbf{3. Concluding remarks}

\bigskip

\noindent In this paper we have established a sequential confidence interval methodology to estimate the dynamic contact angle of a sessile saliva drop. This will help the researchers and practitioners to build an idea about the growth of the pandemic in general or by focusing on specific countries. Since a contact angle has to be measured using a heavy-duty apparatus, a sequential rule also appears to be beneficial by offering a reduction in time and cost. We introduced a new circular model called as the \textit{half-circular wrapped-exponential} distribution to model the angles, which can only spread over half a circle. This new model was seen to fit better than some of the existing ones in the literature. Depending on the width $d$ of the interval fixed by the experimenter, the mean contact angle of the droplet was seen to be between 0.41 and 0.56 radians or 23.49 and 32.08 degrees. On the other hand the drying time of the saliva droplet was seen to be between 61 and 80 seconds.

\bigskip

\noindent\textbf{References}

\bigskip

[1] S. Balusamy, S. Banerjee and K. C. Sahu, Lifetime of sessile saliva droplets in the context of SARS-CoV-2, \textit{Int. J. Heat Mass Trasf.}, 123, 105178 (2021).

[2] R. Mittal, R. Ni, and J.-H. Seo, The flow physics of COVID-19, \textit{J. Fluid Mech.} 894, F2 (2020).

[3] R. Bhardwaj and A. Agrawal, Likelihood of survival of coronavirus in a respiratory droplet deposited on a solid surface, \textit{Phys. Fluids} 32 (6), 061704 (2020).

[4] E. P. Vejerano and L.C. Marr, Physico-chemical characteristics of evaporating respiratory fluid droplets, \textit{J. R. Soc.} Interface 15 (139), 20170939 (2018).

[5] S. Chaudhuri, S. Basu, P. Kabi, V. R. Unni, and A. Saha, Modeling ambient temperature and relative humidity sensitivity of respiratory droplets and their role in determining growth rate of COVID-19 outbreaks, \textit{Phys. Fluids} 32, 063309 (2020).

[6] E. Albert, B. Tegze, Z. Hajnal, D. Z\'amb\'o, D. P. Szekr\'enyes, A. D\'eak, Z. H\'orv\"olgyi and N. Nagy, Robust contact angle determination for needle-in-drop type measurements, \textit{ACS Omega} 4, 18465-18471 (2019).

[7] K. V. Mardia and P. E. Jupp, \textit{Directional statistics}, 2nd Ed. New York: Wiley.

[8] J. S. Rao and A. Sengupta, \textit{Topics in circular statistics}, New York: World Scientific.

[9] A. V. D. Rao and S. V. S. Girija, \textit{Angular statistics}, Boca Raton, CRC Press.

[10] C. Ley and T. Verdebout, \textit{Applied directional statistics}, Boca Raton, CRC Press.

[11] A. Rambli, I. Mohamed, K. Shimizu and N. Ramli, A half-circular distribution on a circle, \textit{Sains Malay.}, 48 (4), 887-892 (2019).

[12] J. S. Rao and T. J. Kozubowski, New families of wrapped distributions for modeling skew circular data, \textit{Comm. in Stat. Theo. and Meth.}, 33 (9), 2059-2074 (2004).

[13] A. Yilmaz and C. Bi\c cer, A new wrapped exponential distribution, \textit{Math. Sci.}, 12, 285-293 (2018).

[14] S. Joshi and K. K. Jose, Wrapped Lindley distribution, \textit{Comm. in Stat. Theo. and Meth.}, 47 (5) 1013-1021 (2018).

[15] S. Banerjee and N. Mukhopadhyay, A general sequential fixed-accuracy confidence interval estimation methodology for a positive parameter: Illustrations using health and safety data, \textit{Ann. of Inst. of Stat. Math.}, 68, 541-571 (2016).

[16] N. Mukhopadhyay and S. Banerjee, Purely sequential and two-stage bounded-length confidence intervals for the Bernoulli parameter with illustrations from health studies and ecology, \textit{P.K. Choudhary et al. (eds.), Ordered Data Analysis, Modeling and Health Research Methods}, Springer Proceedings in Mathematics \& Statistics 149.

[17] S. R. Bapat, On purely sequential estimation of an inverse Gaussian mean, \textit{Metrika}, 81, 1005-1024 (2018a).

[18] S. R. Bapat, Purely sequential fixed accuracy confidence intervals for $P(X > Y)$ under bivariate exponential models, \textit{Am. J. of Math. and Manag. Sci.}, 37, 386-400 (2018b).

[19] A. Khalifeh, E. Mahmoudi and A. Chaturvedi, Sequential fixed-accuracy confidence intervals for the stress-strength reliability parameter for the exponential distribution: two-stage sampling procedure, \textit{Comp. Stat.}, https://doi.org/10.1007/s00180-020-00957-5.

[20] S. Banerjee and N. Mukhopadhyay, A general sequential fixed-accuracy confidence interval estimation methodology for a positive parameter: illustrations using health and safety data, \textit{Ann. of the Inst. of Stat. Math.}, 68, 541-570. (2016).

\end{document}